\documentclass[reprint]{revtex4-1}
\usepackage{graphicx}
\graphicspath{{Fig/}}
\usepackage{dcolumn,placeins}
\usepackage{amsmath}
\usepackage{amssymb}
\usepackage{amsthm}
\usepackage{bm,bbm}

\newcommand{\f}{\frac}
\newcommand{\ep}{\varepsilon}

\newcommand{\E}{\ensuremath{\mathbb{E}}}
\renewcommand{\P}{\ensuremath{\mathbb{P}}}
\newcommand{\xn}{\overline{X}_V}

\newcommand{\Ma}{{\mathcal M}^a}
\newcommand{\ind}{\ensuremath{{1}}}
\newcommand{\diff}{\mathop{}\mathopen{}\mathrm{d}}
\newcommand{\s}{\underset{|u-v| \le \delta}{\sup_{u,v \le t}}}
\newtheorem{proposition}{Proposition}

\newtheorem{theorem}{Theorem}[proposition]
\newtheorem{corollary}{Corollary}[proposition]

\begin{document}

\title{Insights into the variability of nucleated amyloid polymerization by a minimalistic model of stochastic protein assembly} 

\author{Sarah Eug\`ene}
\email{Sarah.Eugene@inria.fr}
\affiliation{\mbox{INRIA, Domaine de Voluceau, Rocquencourt, B.P. 105, 78153 le Chesnay Cedex, France}}
\affiliation{Sorbonne Universit\'es, UPMC Universit\'e Pierre et Marie Curie, UMR 7598, Laboratoire Jacques-Louis Lions, F-75005, Paris, France}

\author{Wei-Feng Xue}
\email{W.F.Xue@kent.ac.uk}
\homepage{\\\href{http://www.kent.ac.uk/bio/profiles/staff/xue.html}{http://www.kent.ac.uk/bio/profiles/staff/xue.html}}
\affiliation{School of Biosciences, University of Kent, Canterbury, Kent, CT2 7NJ, UK}

\author{Philippe Robert}
\email{\mbox{Philippe.Robert@inria.fr, {corresponding author}}}
\homepage{\href{http://team.inria.fr/rap/robert}{http://team.inria.fr/rap/robert}}
\affiliation{\mbox{INRIA, Domaine de Voluceau, Rocquencourt, B.P. 105, F-78153 le Chesnay Cedex, France}}

\author{Marie Doumic}
\email{Marie.Doumic@inria.fr}
\homepage{\\\href{https://www.rocq.inria.fr/bang/Marie-Doumic}{https://www.rocq.inria.fr/bang/Marie-Doumic}\\{PR and MD contributed equally to supervise this work.}}
\affiliation{\mbox{INRIA, Domaine de Voluceau, Rocquencourt, B.P. 105, F-78153 le Chesnay Cedex, France}}
\affiliation{Sorbonne Universit\'es, UPMC Universit\'e Pierre et Marie Curie, UMR 7598, Laboratoire Jacques-Louis Lions, F-75005, Paris, France}

\date{\today}
\begin{abstract}
Self-assembly of proteins into amyloid aggregates is an important biological phenomenon associated with human diseases such as Alzheimer's disease. Amyloid fibrils also have potential applications in nano-engineering of biomaterials. The kinetics of amyloid assembly show an exponential growth phase preceded by a lag phase, variable in duration as seen in bulk experiments and experiments that mimic the small volumes of cells. Here, to investigate the origins and the properties of the observed variability in the lag phase of amyloid assembly currently not accounted for by deterministic nucleation dependent mechanisms, we formulate a new stochastic minimal model that is capable of describing the characteristics of amyloid growth curves despite its simplicity. We then solve the stochastic differential equations of our model and give mathematical proof of a central limit theorem for the sample growth trajectories of the nucleated aggregation process.  These results give an asymptotic description for our simple model, from which closed form analytical results capable of describing and predicting the variability of nucleated amyloid assembly were derived. We also demonstrate the application of our results to inform experiments in a conceptually friendly and clear fashion. Our model offers a new perspective and paves the way for a new and efficient approach on extracting vital information regarding the key initial events of amyloid formation.
\end{abstract}


\maketitle 

\section*{Introduction}

The amyloid conformation of proteins is of increasing concern in our society because they are associated with devastating human diseases such as Alzheimer's disease, Parkinson's disease, Huntington's disease, Prion diseases and type-2 diabetes~\citep{Knowles2, Chiti}. The fibrillar assemblies of amyloid are also of considerable interest in nano-science and engineering due to their distinct functional and materials properties~\cite{Fowler, Schwartz, Knowles3}. Elucidating the molecular mechanism of how proteins polymerize to form amyloid oligomers, aggregates and fibrils is, therefore, a fundamental challenge for current medical and nanomaterials research.

Amyloid diseases are associated with the aggregation and deposition of mis-folded proteins in the amyloid conformation~\cite{Knowles2, Chiti}. Amyloid aggregates form through nucleated polymerization of monomeric protein or peptide precursors (e.g.~\cite{Xue2, Kashchiev, Ferrone, Collins, Knowles4}). The slow nucleation process that initiates the conversion of proteins into their amyloid conformation is followed by exponential growth of amyloid particles, resulting in growth of amyloid fibrils that is accelerated by secondary processes such as fibril fragmentation and aggregate surface catalyzed heterogeneous nucleation~\cite{Radford, Knowles4, Cohen, Xue2} (Figure~\ref{fig:1}). Current mathematical description of protein assembly into amyloid are based on systems of mass-action ordinary differential equations, and they have been successful in describing the average behaviour of amyloid assembly observed by kinetic experiments (e.g.\cite{Radford, Knowles4}). The formation kinetics of amyloid aggregates has been studied extensively by bulk \emph{in vitro} experiments in volumes typically in the range of hundreds of $\mu$L or larger~\cite{Radford}, but has also been observed recently in elegant microfluidic experiments in pL to nL range, more closely mimicking physiological volumes in tissues and cellular compartments~\cite{Knowles5}. Amyloid growth experiments typically follow the appearance of amyloid aggregates or the depletion of monomers as function of time, yielding information regarding the rate of the exponential growth and the length of the lag phase under different protein concentrations at fixed volumes. A hitherto overlooked piece of information that can be derived from these kinetic experiments is the observed variation between experimental repeats, which may hold the key to understanding the early rare nucleation events of amyloid formation~\cite{Radford, Szavits,Eaton,Hofrichter}. However, current deterministic models cannot describe variability, thus, unable to address whether the observed variations in lag phase length reflect subtle experimental differences between the replicates, contributions from the stochastic nature of the nucleation mechanism, or a combination of both factors. As shown recently by Szavits-Nossan and co-workers using a stochastic nucleated growth model, rare nucleation events are expected to dictate the behaviour and variability of amyloid formation in small volumes such as in cellular compartments~\cite{Szavits}. Understanding these rare initial nucleation events of amyloid formation and the variability resulting from the stochastic nature of nucleation, therefore, is of paramount importance in the fundamental understanding of amyloid diseases and in controlling amyloid formation.

\begin{figure}[hbtp]
\centering
\includegraphics[width=8.4cm,height=4.4cm]{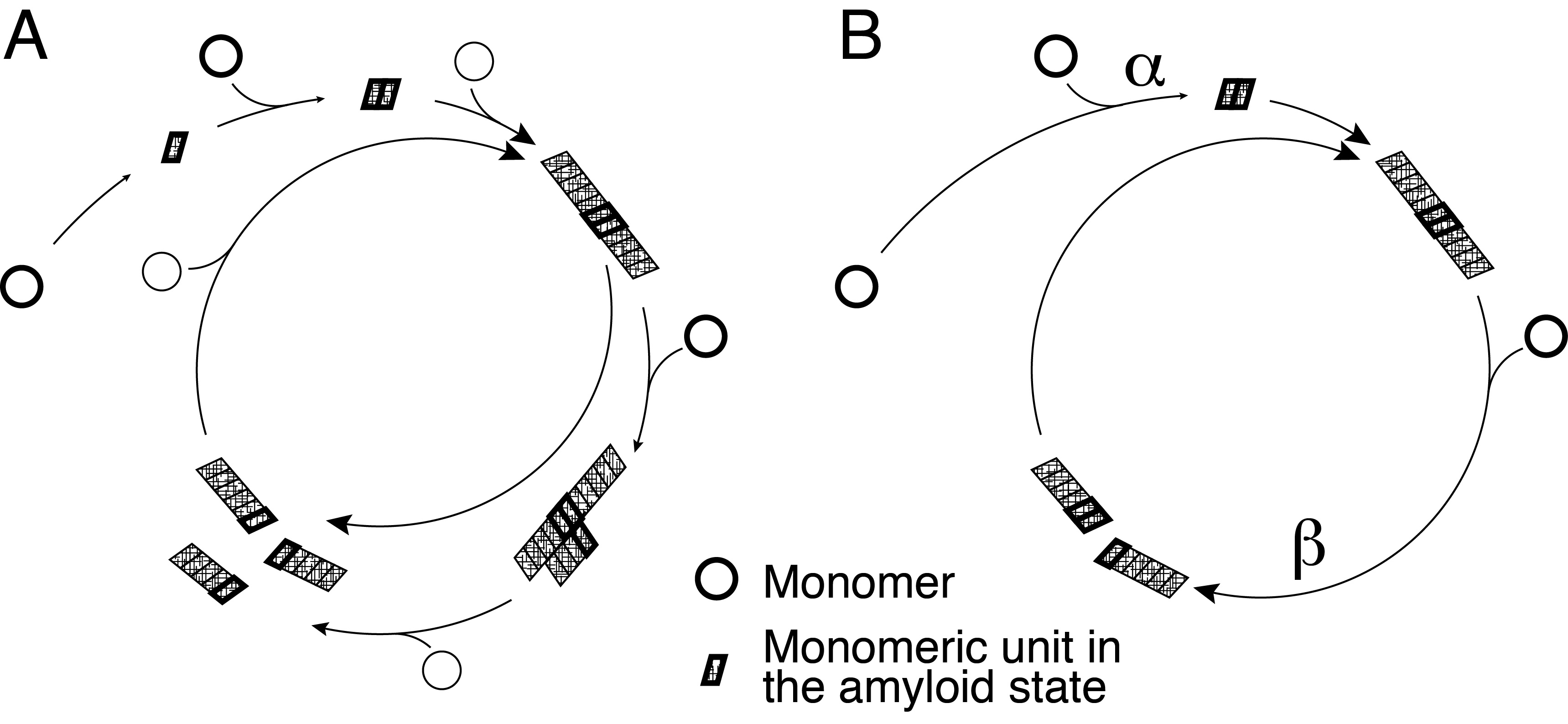}
\caption{(A) represents the complete model, with conformation step, nucleation, and  different possible secondary pathways: polymerisation, lateral polymerisation and fragmentation. 
(B) is our simplified model represented by Reactions~\eqref{reac:alpha} and~\eqref{reac:beta}\label{Fig:SCHEME}: $\alpha$ is the rate of the ignition phase, the take-off and $\beta$ the rate of all possible secondary pathways to the formation of polymers.}\label{Fig1v4.jpg}
\end{figure}

Here, we present a new stochastic protein assembly model with the aim to capture the fundamental features of amyloid self-assembly that includes their stochastic nature, and still allow a fully rigorous mathematical analysis of these processes (Figure~\ref{Fig1v4.jpg}). In this spirit, our model contains minimal possible complexity needed to describe a nucleated protein polymerization process, allowing us to study it theoretically in a mathematically rigorous manner, but still allowing useful comparison to experimental data. From our minimal model, we derive a closed form formula that can describe and predict variability in the lag phase duration of nucleated protein assembly by giving a proof to a central limit theorem for our model. Our results demonstrate how stochasticity at the molecular level may influence the kinetics of the total reaction population at a macroscopic scale depending on the relative rates of nucleation and exponential growth, and on reaction volume. We also show how new information relevant to any specific nucleated amyloid assembly can be gained in a conceptually simple and clear manner by applying our analytical results to the analysis of published $\beta_2 m$ amyloid assembly kinetics data~\cite{Radford}. We demonstrate that our model qualitatively captures key features of the data such as parallel progress of the curves and the order of magnitude for the rates of the self-accelerating reactions. We also show that the intrinsic stochastic nature of nucleation alone cannot explain the observed variability in lag phase length for published $\beta_2 m$ amyloid assembly data acquired in large (100 $\mu$L) volumes suggesting alternative mechanistic assembly steps and additional experimental sources that contribute to the variability in the observed amyloid growth curves. Our approach represents the basis for the development of extensive and tractable stochastic models, which will allow the variability information from amyloid growth kinetics experiments to be used to inform the fundamental molecular mechanisms of the key rare initial events of amyloid formation that may be involved in producing early on-pathway cytotoxic species associated with amyloid disease. 

Supplemental material at [URL] presents the mathematical background of these results, in particular the rigorous proofs of the convergence results, the precise mathematical characterization of the variability of the assembly process and, finally, some simulations of these stochastic processes. 

\section*{A Phenomenological Stochastic Model}\label{FirstSec}
\label{sec:model1}
\begin{figure}[hbtp]
\centering
\hbox to\hsize{\hss\includegraphics{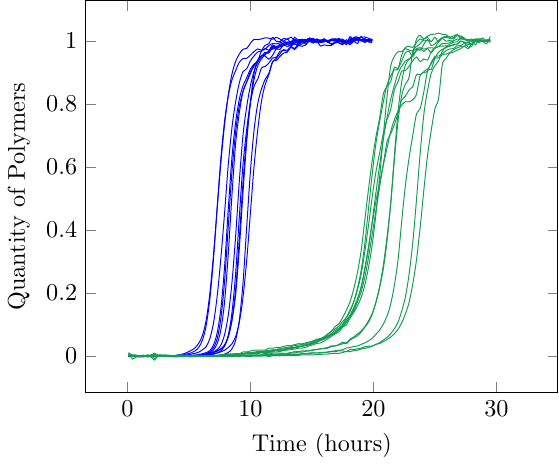}\hss}
\caption{Twelve experimental timecourse of polymerised mass for two given initial concentrations of monomers: $122\hspace{1mm} \mu M$ (blue) and $30.5\hspace{1mm}  \mu M$  (green) published in~\cite{Radford}\label{fig:1}.}
\end{figure}

To make the model as simple as possible, we consider two distinct types of monomers, we call these species monomers and polymerised monomers, respectively. The polymerised monomers represent all monomers in the amyloid conformation in the aggregates. Its amount may be viewed as representing the total polymerised mass, captured for instance by Thioflavine T (ThT) measurements, as in Figure~\ref{fig:1}.
Such a simplification is also justified by the fact that current kinetics measurements of amyloid growth exhibit variability on the timecourse of the total polymerised mass, without giving any information on the size distribution of fibrils. Previous studies (see for instance~\cite{DR}, Supplemental material (S.M.)~2) have shown that  the detail of the reactions of secondary pathways, such as a fragmentation kernel, may have a major impact on  the size distribution of polymers, but comparably smaller effects on the timecourse of the polymerised mass.  

We thus consider two distinct species in our model: monomers, ${\cal  X}$,  and polymerised monomers, ${\cal  Y}$. We then consider $X_V(t)$ and $Y_V(t)$ to be the respective numbers of particles of each species at time $t$ in a fixed volume $V$. Initially, it is assumed that there are only $M$ monomers: $X_V(0)=M$ and $Y_V(0)=0$. We denote $m=M/ (V{\cdot}N_A)$ the initial molar concentration of monomers, where $N_A$ is the Avogadro constant. For convenience in the calculations hereinafter, we introduce the notation $V_A = V{\cdot}N_A$. 

Thus, the chemical reactions associated with this simple model are as follows:
\begin{align}
{\cal X}+{\cal X} &\stackrel{{\alpha}/{V_A^2}}{\underset{}{\longrightarrow}} 2 {\cal Y},
\label{reac:alpha}\\
{\cal X}+{\cal Y} &\stackrel{{\beta}/{V_A^2}}{\underset{}{\longrightarrow}} 2 {\cal Y},\label{reac:beta}
\end{align}
where $\alpha/V_A^2$ and $\beta/V_A^2$ are rates of the reactions with rate constants of $\alpha>0$ and $\beta>0$. These reactions describe the following features of a nucleated polymerisation of proteins that characterises amyloid assembly (see Figure~\ref{Fig:SCHEME} for an explanatory scheme of the reactions):

\begin{enumerate}
\item[---] Reaction~\eqref{reac:alpha}: We call this step "ignition" since it models the starting point of the polymerisation process. Here, we represents this step as the simplest possible concentration dependent nucleation step that converts two monomers into two monomers that are growth competent (equivalent to two polymerised monomers). In our model, this reaction will occur in a stochastic way.  Following the principles of the law of mass action,  the encounter of two chemical species  occurs  at a rate  proportional to the product of the \emph{concentrations} of each species.  Therefore two given monomers disappear to produce two polymerised monomers  at a rate  ${\alpha}/{V^2}$. 
\item[---] Reaction~\eqref{reac:beta}: We call this second step "conversion", which we modelled as a self-accelerating autocatalytic process. Here, given a monomer and a polymerised monomer, the monomer converts into a polymerised monomer at a rate ${\beta}/{V^2}$.  This is representative of a range of accelerating secondary pathway reactions such as fragmentation, lateral growth, and aggregate surface catalyzed second nucleation. In this sense, our model may be viewed as a simplification and amalgamation of several mechanistic models. Even though different secondary processes lead to very different size distributions of fibrils, they affect the total polymerised mass, represented here by the quantity of the species $\cal Y$, in a qualitatively similar way in that they provide acceleration of growth through positive feedback. In particular, we expect our model to behave qualitatively similarly to the mechanistic model  described in~\cite{Szavits}, which includes nucleation, polymerization, and fragmentation as a self-accelerating process.
\end{enumerate}

\noindent
{\bf Stochastic Evolution.} Any given pair of monomers  reacts together by Reaction~\eqref{reac:alpha}  at rate ${\alpha}/{V_A^2}$, whereas for a given pair of monomer/polymerized monomer reacts by Reaction~\eqref{reac:beta} at rate ${\beta}/{V_A^2}$.  Let $M_V$ be  the initial number of monomers and the  random variable describing the number of  monomers remaining  at time $t$ is denoted by $X_V(t)$.  By taking  into account  the  ${X_V(X_V{-}1)}/{2}$  monomers pairs,  and  the $X_V(M_V{-}X_V)$ monomers/polymerised  pairs,  the variable  $X_V(t)$ has  jumps of size $-2$ or $-1$ which occur at the following rates
\begin{equation}\label{eq:X1} 
 X_V \mapsto
\begin{cases}
  X_V{-}2 &\displaystyle \text{ at rate } \frac{X_V(X_V{-}1)}{2}\times \frac{\alpha}{V^2}, \\
 X_V{-}1 & \quad ``\qquad \displaystyle X_V(M_V{-}X_V)  \times \frac{\beta}{V^2}.
\end{cases}
\end{equation}
The conservation of mass gives the additional relation  $X_V(t){+}Y_V(t){=}M_V$. As noticed previously, in the description of Reactions~\eqref{reac:alpha} and~\eqref{reac:beta} above,   this representation  is completely coherent with  the law of mass action.  
\subsection{Asymptotic Evolution of the Number of Monomers}
Assuming that the volume $V$ is large and the initial concentration of monomers remains constant and equal to $m>0$, i.e.  the initial number of monomers $M_V$ is such that ${M_V}/{V_A} \sim m$, we can derive the following:   \\

\noindent
{\bf Polymerisation occurs on the time scale $\mathbf{t{\mapsto}V_A t}$.}\ \
Let $(\overline{X}_V(t))$ be the scaled process defined by
\begin{equation}\label{eq:defXbar}
\overline{X}_V(t) = \frac{X_V(V_At)}{V_A}.
\end{equation}
In Equation~\eqref{eq:defXbar}, the time scale of the process $(X_V(t))$ is accelerated with a factor $V_A$.  As it will be seen, as $V$ gets large,  $t \rightarrow V_A t$ is the correct time scale to observe the decay of $(X_V(t))$ on the space scale proportional to $V_A$. 

Assuming for the moment that $(\overline{X}_V(t))$ is converging in distribution, Relations~\eqref{eq:X1}  then suggest that its limit $(\bar{x}(t))$ should satisfy the following ODE
\begin{equation}\label{eqdiffx}
\frac{\diff \bar{x}}{\diff t}=-\alpha \bar{x}(t)^2 -\beta \bar{x}(t)(m-\bar{x}(t)), \text{ with } \bar{x}(0)=m. 
\end{equation}
The following result shows that this is indeed the case. 
\begin{proposition}[Law of large numbers]
If  the initial number $M_V$ of monomers  is such that
\[
\lim_{V\to+\infty}\frac{M_V}{V_A}=m>0,
\]
then, as $V$ goes to infinity, the process $(\overline{X}_V(t) )$ converges in distribution to $(\bar{x}(t))$, solution of Equation~\eqref{eqdiffx}, given by the formula
\begin{equation}\label{eq:explicit:x}
\bar{x}(t)=m\frac{\beta}{\alpha}\frac{1}{e^{\beta m t} -1 + {\beta}/{\alpha}}.
\end{equation}
\label{prop:LLN}
\end{proposition}
The proof is classical~\cite{EthierKurtz}, we recall it in  Sections~\ref{suppl:1} and~\ref{suppl:analy} of supplemental material, we comment on the relative influence of the parameters $\alpha$ and $\beta$ on the deterministic curve, see supplemental figure~\ref{suppl:fig}.  

In order to be able to quantify the variability of experimental replicates, we need to go further, to a second order approximation. This is given by the following  central limit theorem. 
\begin{proposition}\label{prop:CLT}
If the initial number $M_V$ of monomers is such that
\[
M_V=mV_A+o\left(\sqrt{V_A}\right),
\]
for $m>0$,  then, for the convergence in distribution,
\[
\lim_{V\to+\infty} \left(\frac{X_V(V_At) - V_A \bar{x}(t)}{m\sqrt{V_A}}\right)=(U(t)),
\]
where $U(t)$ is the unique solution of the following stochastic differential equation:
\begin{multline}\label{eq:U}
\diff U(t) =  \frac{\beta \sqrt{ \alpha}\sqrt{e^{\beta m t } +1}}{\alpha e^{ \beta m t} +\beta-\alpha}\,\diff W(t)\\ - \beta m\frac{e^{ \beta m t}{+}1{-}{\beta}/{\alpha}}{e^{ \beta m t}{-}1{+}{\beta}/{\alpha}} U(t)\diff t, 
\end{multline}
 with $U(0)=0$ and $(W(t))$ denotes a standard Brownian motion.
\end{proposition}
The proof to proposition~\ref{prop:CLT} is postponed in Section~\ref{suppl:2} of supplemental material,  together with an explicit formulation and an analysis of the influence of the parameters $\alpha$ and $\beta$ on the stochasticity of the reactions. We found that the smaller the ratio ${\alpha}/{\beta}$ is, the more important the influence of the stochasticity on the lag-time, but the less important for the following of the reaction. 
This is quantified in the following study of the stochastic time for $\delta$ completion below. 

\begin{figure}[hbtp]
\centering

\hbox to\hsize{\hss\includegraphics{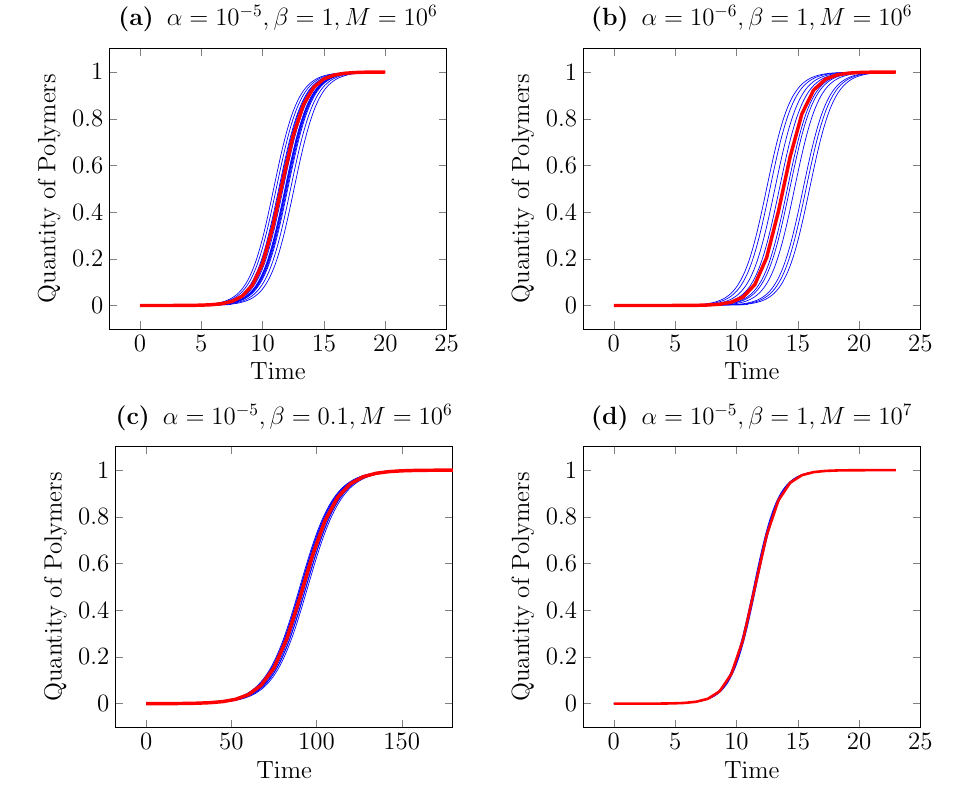}\hss}

\caption{Simulations of the model with different parameters. The red curve is the first order obtained in Proposition 1.}
\label{Fig:Simus}
\end{figure}

\subsection{Asymptotics of the Time for $\bm{\delta}$ Reaction Completion}
To quantify the effect of $\alpha$ and $\beta$ on the stochasticity of the reactions, we define  the {\em time for $\delta$ reaction completion}, where $0<\delta<1$ is a percentage, as the following stopping time
\[
 T_V(\delta) = \inf \{ t >0, X_V(t) \le (1- \delta) M_V\} \\
\]
where $T_V$ is the first time when there is a $\delta$ fraction of polymers is produced. $T_V$ for $\delta$ small - $5$ to $10\%$ - represents an alternative definition for the lag-time of the reaction\cite{DR}. 

\begin{figure}[hbtp]
\hbox to\hsize{\hss\includegraphics{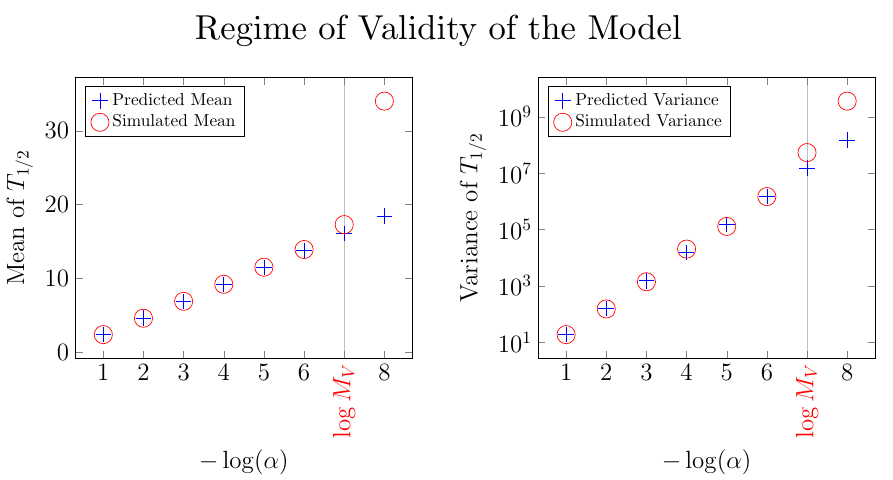}\hss}
\caption{Comparison between the simulations and the predictions to see the regime of parameters where the calculations are valid. For these simulations, we fixed $M_V = 10^7$ monomers, $\beta=1$, and made $\alpha$ varying.
\label{Fig:RegPar} } 
\end{figure}

The following theorem gives a central limit result for $T_V(\delta)$ as $V$ goes to infinity. Note that due to the change in the time scale, we need to rescale $T_V$ by $V$ to get a limit.
\begin{theorem}[Asymptotics of the time for degree of reaction completion $\delta$]\label{lagtheo}
If the initial number $M_V$ of monomers is such that
\[
M_V=mV_A+o\left(\sqrt{V_A}\right),
\]
for $m>0$  then, for the convergence in distribution
\begin{enumerate}
\item {\bf Law of Large Numbers}.
\begin{equation}\label{tdelta}
\lim_{V\to+\infty} \frac{T_V(\delta)}{V_A}  =t_\delta \stackrel{\text{def.}}{=}  \frac{1}{\beta m} \log\left(1 + \frac{\beta \delta}{ \alpha (1-\delta)}\right). 
\end{equation}
\item {\bf Central Limit Theorem}.
\[
\lim_{V\to+\infty}  \frac{T_V(\delta) - V_A t_\delta}{\sqrt{V_A}}=\frac{U(t_\delta)}{m[\alpha(1-\delta)^2+\beta\delta(1-\delta)]}
\]
\end{enumerate}
where  $(U(t))$ is the solution of the SDE~\eqref{eq:U}. 
\end{theorem}
The proof of Theorem~\ref{lagtheo} is given in Section~\ref{suppl:3} of supplemental material. 
Supplemental Figure~\ref{suppl:Thalf1} illustrates and  the law of large numbers and the central limit theorem for $T_{1/2}$.
Note that the definition of $t_\delta$, which is the limit of the stochastic times $T_V(\delta)/V$ when $V\rightarrow \infty$ is coherent with the definition of the \emph{deterministic}  time of $\delta$ reaction completion as
$$t_\delta=\inf\{t>0,\,\bar x (t) \leq (1-\delta) m\}=\bar{x}^{-1}((1-\delta)m),$$
where $(\bar{x}(t))$ is given by Equation~\eqref{eq:explicit:x}.
Thus, for any given experiment, the distance between a realization of $T_V(\delta)/V_A$ and $t_\delta$ is being given by the explicit formula above. We can therefore derive its stochastic behaviour. The following corollary establishes its variance.
\begin{corollary}[Variance of  Time $T_V(\delta)$]\label{corolVar}
Under the assumptions of the above theorem and with its notations, the variance $\sigma_V^2$ of the time for $\delta$ completion has a limit $\sigma^2$, when $\alpha\ll \beta$ 
\begin{equation}\label{Vardelta}
\lim_{V\to+\infty} \sigma_V = \sigma \sim{ \frac{\sqrt{3}}{\sqrt{2}m\sqrt{M_V\alpha\beta}}}.
\end{equation}
\end{corollary} 
The proof of corollary~\ref{corolVar} is given in Section~\ref{suppl:4} of supplemental material, together with the exact formula for $\sigma$. It should be noted that this representation of $\sigma$ is independent of $\delta$. This suggests that the fluctuations do not depend on $\delta$, and therefore, the growth curves predicted by our simple model are all parallel for any given concentration.
Figures~\ref{Fig:Fit} (c) and~~\ref{Fig:Fit} (d) below have been obtained by centering the 12 curves of Figures~\ref{Fig:Fit} (a) and~\ref{Fig:Fit} (b) at the half-time corresponding to $\delta=1/2$. As it can be seen, the  times $T_V(\delta)$ for  $0.4\leq \delta\leq \ 0.7$  are then also  superimposed: the curves are identical for this range of values.  This is an illustration of the above relation~\eqref{Vardelta}. The exact mathematical formulation of this phenomenon is shown in supplemental material, Proposition~\ref{suppl:para} of Section~\ref{suppl:4}. Simple as it is, our model captures well this feature experimentally observed. Also, it emphasizes the fact that we can take different values for $\delta$ without having an influence on  the study. A difficulty however lies in the fact that when the numerical values of the constant $\alpha$ above is in the order of $1/M_V$, then the convergence itself may be a problem, as it can be seen on Figure~\ref{Fig:RegPar}.

\subsection{Estimation of the parameters}
In this section, we tested the results obtained with our minimalistic stochastic model on the data published in~\cite{Radford}. In these experiments, there are 12 replicate kinetic traces  reported for each sample concentration in constant $100\mu L$ reaction volume. The parameters $\alpha$ and $\beta$ are obtained by fitting the mean half-time $t_{1/2}$ and the mean slope $k$ of  the curves at $t_{1/2}$. More precisely, using Formula~\eqref{eqdiffx} for $k$ and Relation~\eqref{tdelta} for $t_{1/2}$, gives
\[
t_{1/2}=\log\left(1+{\beta}/{\alpha}\right)/{\beta m}\text{ and }
k={m\beta}\left(1+{\beta}/{\alpha}\right)/{4}
\]

\begin{figure}[hbtp]
\centering

\hbox to\hsize{\hss\includegraphics{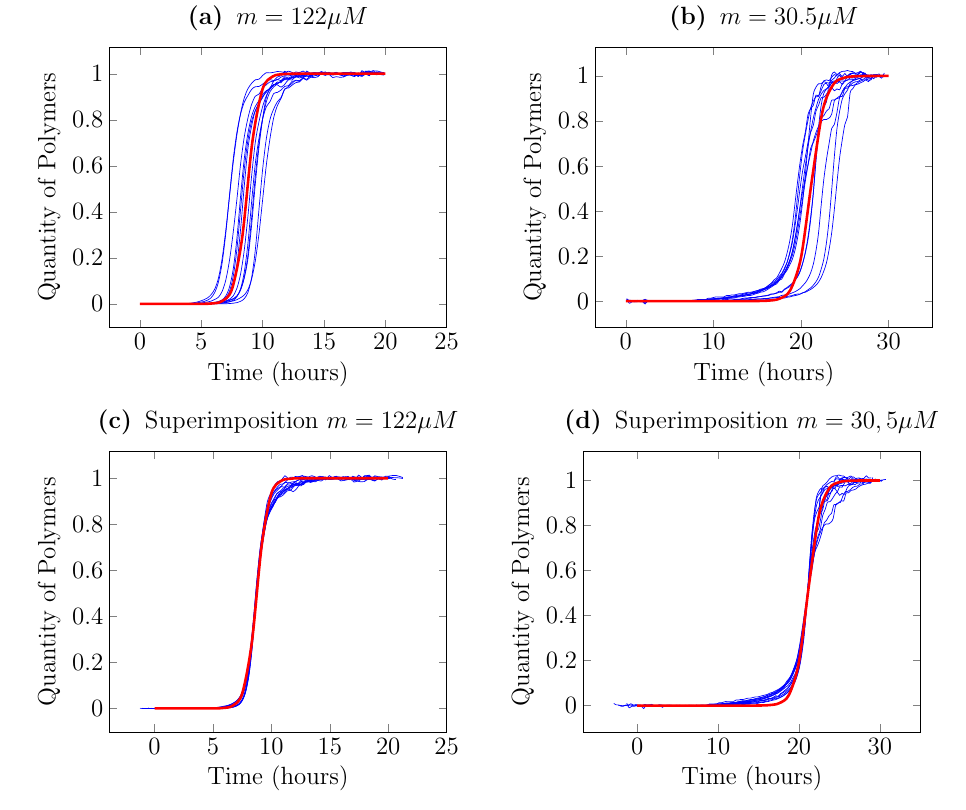}\hss}

\caption{(a) and (b): Experimental timecourse of polymerised mass for 12 different experiments. (c) and (d): with a centering at the $T_V(1/2)$ of each curve. Published in~\cite{Radford}. The red curve is the predicted mean with the estimated parameters.}\label{Fig:Fit}
\end{figure}

Table~\ref{tab:1} shows a summary of our analysis. The constants $\alpha$ and $\beta$ in $h^{-1}.\mu M^{-1}$, and the variance in $h$ are shown for each of the concentration used. The superposition of the experimental curves around the predicted mean is illustrated on Figure~\ref{Fig:Fit}, Figures~\ref{Fig:Fit} (c) and~\ref{Fig:Fit} (d) have been obtained by centering the 12 curves of Figures~\ref{Fig:Fit} (a) and~\ref{Fig:Fit} (b) at the half-time corresponding to $\delta = 1/2$. As can be seen, the curves can be superimposed for $0.4\le \delta \le 0.7$. This is consistent with the relation~\ref{Vardelta}. Our analysis further demonstrates two important insights. 

\noindent
\begin{table*}
\scalebox{1}{
\begin{tabular}{|c||c|c|c|c|}
\hline
$\bf {m (10^{-6} M)}$ & $\bf {\alpha (h^{-1}.M^{-1})}$ & $\bf{\beta (h^{-1}.M^{-1})}$ & \bf{Experimental Std $(h)$} & \bf{Predicted Std $(h)$}  \\ \hline\hline
$12.3  $&$  6.18{\cdot}10^{-7} $&$  5.07{\cdot}10^{4} $&$  7.95      $&$  5.34{\cdot}10^{-2} $\\ \hline
$14.6  $&$  2.81{\cdot}10^{-6} $&$  4.54{\cdot}10^{4} $&$  2.98      $&$  2.05{\cdot}10^{-2} $\\ \hline
$16.7  $&$  1.59{\cdot}10^{-4} $&$  3.75{\cdot}10^{4} $&$  2.68      $&$  2.45{\cdot}10^{-3} $\\ \hline
$17.0  $&$  1.88{\cdot}10^{-3}  $&$  3.70{\cdot}10^{4} $&$  1.52      $&$  6.98{\cdot}10^{-4} $\\ \hline
$29.5  $&$  1.40{\cdot}10^{-5}  $&$  3.34{\cdot}10^{4} $&$  2.13      $&$  3.7{\cdot}10^{-3}  $\\ \hline
$30.2  $&$  2.89{\cdot}10^{-2}  $&$  2.96{\cdot}10^{4} $&$  2.57      $&$  8.40{\cdot}10^{-5} $\\ \hline
$30.5  $&$  9.57{\cdot}10^{-8} $&$  4.16{\cdot}10^{4} $&$  1.53      $&$  3.84{\cdot}10^{-2} $\\ \hline
$43.7  $&$  7.99{\cdot}10^{-3}  $&$  2.35{\cdot}10^{4} $&$  2.10      $&$  1.03{\cdot}10^{-4} $\\ \hline
$48.5  $&$  1.68{\cdot}10^{-2}  $&$  2.01{\cdot}10^{4} $&$  1.56      $&$  6.55{\cdot}10^{-5} $\\ \hline
$61.0   $&$  2.61{\cdot}10^{-2}  $&$  2.04{\cdot}10^{4} $&$  1.03      $&$  3.71{\cdot}10^{-5} $\\ \hline
$61.0   $&$  2.22{\cdot}10^{-5} $&$  2.56{\cdot}10^{4} $&$  2.55      $&$  1.14{\cdot}10^{-3} $\\ \hline
$84.1  $&$  4.53{\cdot}10^{-4} $&$  2.24{\cdot}10^{4} $&$  1.59      $&$  1.66{\cdot}10^{-4} $\\ \hline
$102.2 $&$  1.52{\cdot}10^{-3}  $&$  1.88{\cdot}10^{4} $&$  0.62      $&$  7.39{\cdot}10^{-5} $\\ \hline
$122   $&$  1.33{\cdot}10^{-4} $&$  1.75{\cdot}10^{4} $&$  0.90      $&$  1.98{\cdot}10^{-4} $\\ \hline
$123.5 $&$  2.13{\cdot}10^{-4} $&$  1.79{\cdot}10^{4} $&$  0.90      $&$  1.52{\cdot}10^{-4} $\\ \hline
$142.1 $&$  2.58{\cdot}10^{-4} $&$  1.74{\cdot}10^{4} $&$  1.11      $&$  1.13{\cdot}10^{-4} $\\ \hline
$243.5 $&$  1.75{\cdot}10^{-3}  $&$  1.09{\cdot}10^{4} $&$  0.60      $&$  2.46{\cdot}10^{-5} $\\ \hline
\end{tabular}}
\caption{\label{tab:1}Parameters estimated from experimental data published in~\cite{Radford} using our model. The two first columns are the estimated parameters $\alpha$ and $\beta$ from the model. The third column is the experimental standard deviation of $T_V(1/2)$, while the fourth is the standard deviation predicted by our mathematical results for the model with the estimated parameters.  {We see that the estimation for $\beta$ is quite robust, in contrast with that of $\alpha.$}}
\end{table*}
\bigskip

\noindent
Firstly,  we obtained a  more  well-estimated $\beta$ parameter. It is remarkable that the numerical value of $\beta$, which quantifies the conversion step in our model, does not change much for the 15 concentrations tested in the experiments, considering the simplicity of our model. This is not the case for $\alpha$,which quantifies the ignition phase, varies between  $10^{-7}$ and $10^{-13}$.  Here, the parameter $\alpha$ which quantifies the take-off phase (remember that the slope of $(\bar{x}(t))$ at $0$ is $-\alpha m^2$)  is intrinsically estimated with less precision than $\beta$, see Section~\ref{suppl:analy} of supplemental material. This is a limitation of this simple model, and it also reflect the lack of information content in the kinetics data during the lag phase compared to the growth phase.
 

Secondly, {despite good agreement between our analysis and the data in terms of the shapes of the growth curves, the analysis results in a much smaller order of magnitude for the variability among curves compared with experimental data. }  Since the relation $\alpha\ll \beta$ holds in the numerical estimations, Equation~\eqref{Vardelta} gives the approximation $\sigma^2\sim {3}/({2 M_Vm^2\alpha\beta})$ for the variance of the characteristic times of the kinetic traces.  A variance of the order of magnitude observed in the experiments~\cite{Radford} would be obtained by our model for an initial number of monomers $M_V$ in the order of $10^6$. As the number $M_V$ in the experiments of~\cite{Radford} performed in $100 \mu L$ volumes is closer to $10^{15}$, our analysis suggest that the variability observed result from more than a simple stochastic homogeneous nucleation of monomers. This result is consistent with the mechanistic approach used by Szavits-Nossan and co-workers~~\cite{Szavits}, where the authors used a stochastic nucleation-polymerization-fragmentation based model. Thus, our model and analysis of the variance suggest alternative initial rare assembly steps that involve additional complexities such as conformational exchange, and/or additional experimental sources that contribute to the variability in the observed amyloid growth curves.

\bigskip

\section*{Conclusion}
{In this study, we described an approach that represents the molecular mechanisms of amyloid growth in a condensed way to enable the development of a rigorous framework that can describe the stochastic behaviour in addition to the general features of the kinetics of assembly. We adopted a reductionist approach by including minimal complexity using two simple processes, ignition and conversion, to reflect the idea that a protein polymerization reaction accelerated by secondary processes is dictated by the first nucleation event~\cite{Knowles5, Szavits}. With our minimal model, we were able to derive an exact analytical formula for the expected variability among curves as function of relative rates of the ignition and conversion processes, and the reaction volume. This is useful both in exploring the interplay between the reaction rates of nucleation and growth, and the variability in reaction traces, as well as in the analyses of experimental kinetics data. We also see that the stochasticity influences mainly the ignition step: once the reaction accelerates in the conversion step, all curves become parallel and deterministic, as illustrated both by experiments and the model we presented here. Thus, simple as it is, our model captures well the features experimentally observed for amyloid growth curves. Also, it confirms, as expected, that we can take different characteristic times (such as lag time, or growth mid point) when analysing kinetic growth curves. Our model further informed the need for new mechanistic steps or experimental interpretation of the large observed variations in the lag time lengths. Thus, the variation seen in the kinetic traces must be taken into account in addition to the concentration dependent behaviour of the kinetic traces in evaluating and developing mechanistic understanding of amyloid protein assembly processes.

While our model design was not aimed at describing the reality of any specific amyloid forming system with all of their individual associated complexities, our design by pursuing maximum simplicity are complementary to mechanistic approaches such as in~\cite{Radford, Knowles4, Szavits} in capturing global properties of amyloid assembly. Thus, our method allows for a rigorous theoretical treatment and understanding, and provides a basis for model selection on stochastic 'minimal models', each of these models being the condensation of a family of possible stochastic mechanistic models, closer to reality but for which analytical formulae are out of reach.}

\begin{acknowledgments}
{M. Doumic and S. Eug\`ene's research
was supported by ERC Starting Grant SKIPPER$^{AD}$ No. 306321.}
\end{acknowledgments}

\newpage \setcounter{section}{0}\begin{widetext} \newpage 
g
\section*{Supplemental Material: Proofs of the main results and analysis of the parameters}
 
Recall that it is assumed that the parameter $M_V$ is asymptotically proportional to $m$
\[
\lim_{V\to+\infty}\frac{M_V}{V_A}=m.
\]
\subsection{Proof of Proposition~\ref{prop:LLN} (Law of large numbers)}\label{suppl:1}

The proof relies on classical methods of stochastic calculus, see for instance Darling and Norris~\cite{Darling} or  Ethier and Kurtz~\cite{EthierKurtz}. Here, we give a summary of the proof for the completeness.

\noindent
{\bf Stochastic Equations.}  
Let  $\mathcal{N}_{\alpha/V_A^2}(dt)$ [resp.  $\mathcal{N}_{\beta/V_A^2}(dt)$] be a Poisson process with  parameter ${\alpha}/{V_A^2}$ [resp. ${\beta}/{V_A^2}$],  then Relations~\eqref{eq:X1} give that  the random variable  $X_V(t)$ can be represented as a solution of the following stochastic differential equation
\begin{equation}\label{eq:X} 
\diff X_V(\diff t) = - 2 \sum\limits_{i=1}^{{X_V(X_V-1)(s{-})}/{2}} \mathcal{N}_{{\alpha}/{V^2}}^i(\diff t)- \sum\limits_{i=1}^{{\small{X_V(M-X_V)(s{-})}}} \mathcal{N}_{{\beta}/{V^2}}^i(\diff t), 
\end{equation}
with  $X_V(0)=M$ and $f(s{-})$ denotes the limit on the left of $f$ at $s$.  For more  discussion and  results on related models, see  for example  Anderson and Kurtz~\cite{AndersonKurtz} and Higham~\cite{Higham} and references therein. 

By using Equation~\eqref{eq:X}  and $\overline{X}_V$ defined by~\eqref{eq:defXbar}, one gets
\begin{equation}\label{eqaux}
\overline{X}_V(t)=\frac{X_V(V_At)}{V_A} = \f{M_V}{V_A} + \Ma_V(t) - \alpha \int_0^{t} \overline{X}_V\left(\overline{X}_V-\f{1}{V_A}\right)(s)\diff s  - \beta \int_0^t \overline{X}_V\left(\frac{M_V}{V_A}-\overline{X}_V\right)(s)\diff s,
\end{equation}
where $(\Ma_V(t))$ is the martingale
\begin{align*}
\Ma_V(t) =  &- \f{2}{V_A}  \sum_{i=1}^{\infty}  \int_0^{V_At} \ind_{\{i\le X_V(X_V-1)(s{-})/2\}} \left(\mathcal{N}_{{\alpha}/{V^2}}^i(\diff s) - \frac{\alpha}{V_A^2} \diff s \right) \\&- \sum_{i=1}^{\infty} \frac{1}{V_A}\int_0^{V_At} \ind_{\{i\le X_V(M - X_V)(s{-})\}} \left(\mathcal{N}_{{\beta}/{V_A^2}}^i(\diff s) - \frac{\beta}{V_A^2} \diff s \right). 
\end{align*}
Its quadratic variation is given by 
\[
\langle \Ma_V \rangle(t) =  \frac{2\alpha }{V_A} \int_0^t \overline{X}_V\left(\overline{X}_V-\f{1}{V_A}\right)(s) \diff s + \frac{\beta}{V_A} \int_0^t \overline{X}_V\left(\frac{M_V}{V_A}-\overline{X}_V\right)(s) \diff s
\leq \frac{1}{V_A}Ct,
\]
for some constant $C$ since $X_V$ is bounded by $M_V$.  Doob's inequality gives that, with high probability, the martingale $(\Ma_V(t))$ vanishes uniformly on finite intervals: for $\varepsilon>0$,
\[
\P\left( \sup_{0\le s \le t} \left | {\Ma_V(s)} \right|\geq \varepsilon  \right)\le\f{1}{\varepsilon^2}\mathbb{E} \left( \left \langle\Ma_V\right\rangle(t) \right)\leq \frac{1}{ V_A}\f{Ct}{\varepsilon^2}.
\]
We can now show that the sequence $(\xn)_V$ is tight.  Let $$ w_V(\delta) = \underset {u,v \le t} {\sup_{|u-v| \le \delta}} \left | \xn(u) -\xn(v) \right |.$$
Then, Equation~\eqref{eqaux} gives 
\[
w_V(\delta) \le \underset {u,v \le t} {\sup_{|u-v| \le \delta}} \left | \Ma_V(u) -\Ma_V(v) \right | + \delta(\alpha  + \beta)\left(\frac{M_V}{V_A}\right)^2.
\]
Therefore, for $\varepsilon>0$ and $\eta>0$, there exist $\delta_0$ and $V_0$ such that if $\delta\leq \delta_0$ and $V\geq V_0$  then
$\P(w_V(\delta) \ge \varepsilon) \leq \eta. $
Consequently, the sequence $(\overline{X}_V(t))$ is tight, see Ethier and Kurtz~\cite{EthierKurtz} for example. Let $(x(t))$ be one of the  limiting points of $(\overline{X}_V(t))$, it necessarily satisfies the following differential equation
\[
\dot{x} = - \alpha x^2 - \beta x(m-x) \text{ with } x(0)=m. 
\]
\subsection{Proof of Proposition~\ref{prop:CLT} (Central limit theorem)}\label{suppl:2}
With Equation~\eqref{eqaux}, one gets
\begin{multline}\label{eqaux2}
U_V(t) = \frac{X_V(V_At) - V_A x(t)}{m\sqrt{V_A}} 
= \frac{\sqrt{V}\Ma_V(t)}{m} -  \alpha \int_0^t U_V(s)  \left(\xn(s) + x(s)\right) \diff s\\-\beta m \int_0^t U_V(s) \diff s+ \beta  \int_0^t U_V(s) (\xn(s) + x(s))\diff s +\frac{\alpha}{\sqrt{V}} \int_0^t \xn(s) \diff s .
\end{multline}
First note that the process associated to the last term of this expression converges in distribution to zero.
Concerning the martingale term of this relation,  one has 
 \begin{align*}
\left\langle \sqrt{V_A}\frac{\Ma_V}{m} \right\rangle(t) &=  \frac{1}{m^2 } \left[ 2 \alpha \int_{0}^t (\xn){(\xn-1)}(u)  \diff u + \beta \int_0^t \xn(m -\xn)(u) \diff u \right].
\end{align*}
The law of large numbers which has just been proved gives that this process converges to
\[
2\alpha \int_0^t \f{x^2}{m^2}\diff s +\f{\beta}{m^2} \int_0^t x(s)(m-x(s))\diff s=\f{\alpha}{m^2} \int_0^t x(s)^2 \diff s+\frac{1}{m^2}(m-x(t))=\psi(t).
\]
Thus, we get from Theorem~1.4 page~339 of Ethier and Kurtz~\cite{EthierKurtz} that,  as $V$ goes to infinity,  the process $({\sqrt{V}\Ma_V(t)}/{m})$ converges in distribution to 
\[
\int_0^t \sqrt{\dot{\psi}(s)} \diff W(s),
\]
where $(W(t))$ is the standard Brownian motion.

We now prove that the sequence of processes $(U_V(t))$ is tight. 
Let 
$$ w_V(\delta) = \underset {u,v \le t} {\sup_{|u-v| \le \delta}} \left | U_V(u) -U_V(v) \right |,$$
then, by using Equation~\eqref{eqaux2}, one gets
\begin{equation}\label{eqaux3}
w_V(\delta) \le  \underset {u,v \le t}{\sup_{|u-v| \le \delta}}\left | \frac{\sqrt{V_A}\Ma_V(u)}{m} - \frac{\sqrt{V}\Ma_V(v)}{m} \right |+ \s   \frac{\alpha}{\sqrt{V}} \left| \int_u^v \xn(s) \diff s \right|+C_0\s \int_u^v |U_V(s)| \diff s,
\end{equation}
for some  fixed constant $C_0$.  Consequently, 
\[
\underset{s \le t}{\sup} |U_V(s)|  \le \underset{s \le t}{\sup} \left( \left | \frac{\sqrt{V}\Ma_V(s)}{m} \right |\right) +\frac{\alpha}{\sqrt{V}} m t + C_0\int_0^t \underset{\tau \le s}{\sup} |U_V(\tau)| \diff s,
\]
by using Gronwall's lemma, one gets
\[
 \underset{s \le t}{\sup} |U_V(s)| \le  \left[\underset{s \le t}{\sup}\left(\left | \frac{\sqrt{V}\Ma_V(s)}{m}  \right | \right)+\frac{\alpha}{\sqrt{V_A}} m t \right ]e^{C_0 t}.
\]
The convergence of the processes $\left(\sqrt{V}\Ma_V(t)\right)$ shows that the left-hand side of the above expression is bounded with high probability. Relation~\eqref{eqaux3}  and the tightness of $\left(\sqrt{V}\Ma_V(t)\right)$ give then directly the tightness of $(U_V(t))$. 

Let $U$ be a limiting point of the sequence $(U_V(t))$ when $V$ goes to infinity.
Relation~\eqref{eqaux2} shows that $U$ must satisfy the following stochastic  equation
\begin{equation}\label{eqaux4}
 U(t) =  \int_0^t b(s) \diff W(s) +  \int_0^t a(s)U(s) \diff s,
\end{equation}
where 
\begin{equation}\label{ab}
b(t)=\sqrt{\dot{\psi}(t)}= \frac{\beta \sqrt{ \alpha } \sqrt{ e^{\beta m t } +1}}{\alpha e^{\beta m t} + \beta - \alpha}, \qquad a(t)=\beta m  \frac{\beta - \alpha -\alpha e^{\beta m t }}{\beta - \alpha +\alpha e^{\beta m t }} .
\end{equation}
This proves that the process $(U_V(t))$ converges in distribution to $(U(t))$. 
\subsection*{Explicit solution of the SDE for $\mathbf{U}$}
\begin{corollary}
The SDE for $U$ has an explicit solution:
\begin{equation}\label{eq:explicit:U}
U(t) = \frac{e^{\beta m t}}{\left({\beta}/{\alpha} -1 + e^{\beta m t}\right)^2} \int_0^t \frac{\beta }{\sqrt{\alpha}} \left[ \left(\frac{\beta }{\alpha}-1\right) e^{- \beta m s /2} + e^{\beta m s / 2} \right]\left[ \sqrt{1 + e^{-\beta m s}}\right] \diff W_s.
\end{equation}
\end{corollary}
Straightforward stochastic calculus shows that  the right-hand side of Equation~\eqref{eq:explicit:U}  satisfies the Stochastic Differential  Equation associated to Relation~\eqref{eqaux4}. 

\subsection{Proof of Theorem~\ref{lagtheo} (Time for $\bm{\delta}$ reaction completion asymptotics)}\label{suppl:3}
It is enough to prove the central limit theorem. 
\noindent
For $w\geq 0$, since  $\{T_V(\delta)\leq w\}=\{X_V(w)\leq (1-\delta)M_V\}$,
\[
\left\{ \frac{T_V(\delta) - V_A t_\delta}{\sqrt{V_A}}\leq w\right\}=
\left\{X_V\left[V_A(t_\delta+w/\sqrt{V_A})\right]\leq (1-\delta)M_V\right\},
\]
the probability of the event can therefore be expressed as 
\[
\P\left( \frac{X_V\left[V_A(t_\delta{+}w/\sqrt{V})\right]{-}V_A\bar{x}\left(t_\delta{+}w/\sqrt{V_A}\right)}{\sqrt{V_A}}{\leq}\sqrt{V_A}\left(\bar{x}(t_\delta){-}x\left(t_\delta{+}w/\sqrt{V_A}\right)\right){+}o(1)\right).
\]
Hence, by Proposition~\ref{prop:CLT}, for the convergence in distribution
\[
\lim_{V\to+\infty} \frac{X_V\left[V_A(t_\delta+w/\sqrt{V_A})\right]-V\bar{x}(t_\delta+w/\sqrt{V_A})}{m\sqrt{V_A}}=U(t_\delta),
\]
consequently, one gets the convergence
\[
\lim_{V\to+\infty} \P\left( \frac{T_V(\delta) - V_A t_\delta}{\sqrt{V_A}}\leq w\right)=
\P\left(U(t_\delta)\leq \frac{-\dot{\bar{x}}(t_\delta)}{m}w\right)=
\P\left(\frac{U(t_\delta)}{m[\alpha(1-\delta)^2+\beta\delta(1-\delta)]}\le w \right)
\]
The result is proved. 
\subsection{Proof of Corollary~\ref{corolVar} (Variance of the time for $\bm{\delta}$ reaction completion)}\label{suppl:4}
This is a direct consequence of the above central limit theorem and  of Relation~\eqref{eq:explicit:U} and the fact that
\begin{align}
\sigma^2&:=\lim_{V\to+\infty} \E\left[\left(\frac{T_V(\delta) - V_A t_\delta}{\sqrt{V_A}}\right)^2\right]\notag\\
&= \lim_{V\to+\infty} \E\left[\left(\frac{U(t_\delta)}{m[\alpha(1-\delta)^2+\beta\delta(1-\delta)]}\right)^2\right]\notag\\
& =\frac{\alpha}{ m^3\beta^2} \Bigg[ \left( \frac{\beta}{\alpha} -1\right)^2 \frac{1}{2 \beta}\left( 1 - \frac{1}{\left( 1 +{\beta \delta}/{(\alpha(1 - \delta))}\right)^2}\right)+\left(\frac{\beta }{ \alpha}-1\right)\left(\frac{\beta }{ \alpha}+1\right)\frac{\delta}{\beta\delta+\alpha(1-\delta)}\notag\\
&\hspace{6cm}+\left(2\frac{\beta }{ \alpha}-1\right) \frac{1}{\beta} \log\left(1 + \frac{\beta \delta}{ \alpha (1-\delta)}\right)+\frac{1}{\alpha}\frac{\delta}{1-\delta}\rule{0mm}{4mm}\Bigg].\label{eqq}
\end{align}

One can get a more precise result by using the fact that $(U(t))$ is a Gaussian process, by Equation~\eqref{eq:explicit:U} for example, the following representation holds for the $T_V(\delta)$. 
\begin{proposition}\label{suppl:para}
Provided that $\alpha{\ll}\beta$ then, for the convergence in distribution,  as $V$ gets large
\[
T_V(\delta) \sim V_A t_\delta+  \sqrt{V_A} \mathcal{N}\left(0,\frac{3}{ 2 M_V \alpha \beta m^2}\right),
\]
where ${\cal N}(0,x)$ is a center Gaussian random variable with variance $x$. 
\end{proposition}
We illustrate this proposition on Figure~\ref{suppl:Thalf1} below.
This expansion shows that the stochastic fluctuations, the term associated with $\sqrt{V}$, do not depend on $\delta$.  This remarkable property is also true in the experiments: the curves superimpose very well. See Figure~\ref{Fig:Fit} (c) and (d). 
\begin{proof}
The central limit theorem gives
\[
\frac{T_V(\delta) -V_A t_\delta}{ \sqrt{V}}\sim  \frac{U(t_\delta)}{m[\alpha(1-\delta)^2+\beta\delta(1-\delta)] }\sim {\cal N}(0,\sigma^2),
\]
by Equation~\eqref{eq:explicit:U}, where $\sigma$ is defined above. The result follows by using the fact that $\alpha\ll \beta$ in the explicit expression~\eqref{eqq} of $\sigma$. 
\end{proof}
\begin{figure}[hbtp]
\centering
\hbox to\hsize{\hss\includegraphics[scale=0.25]{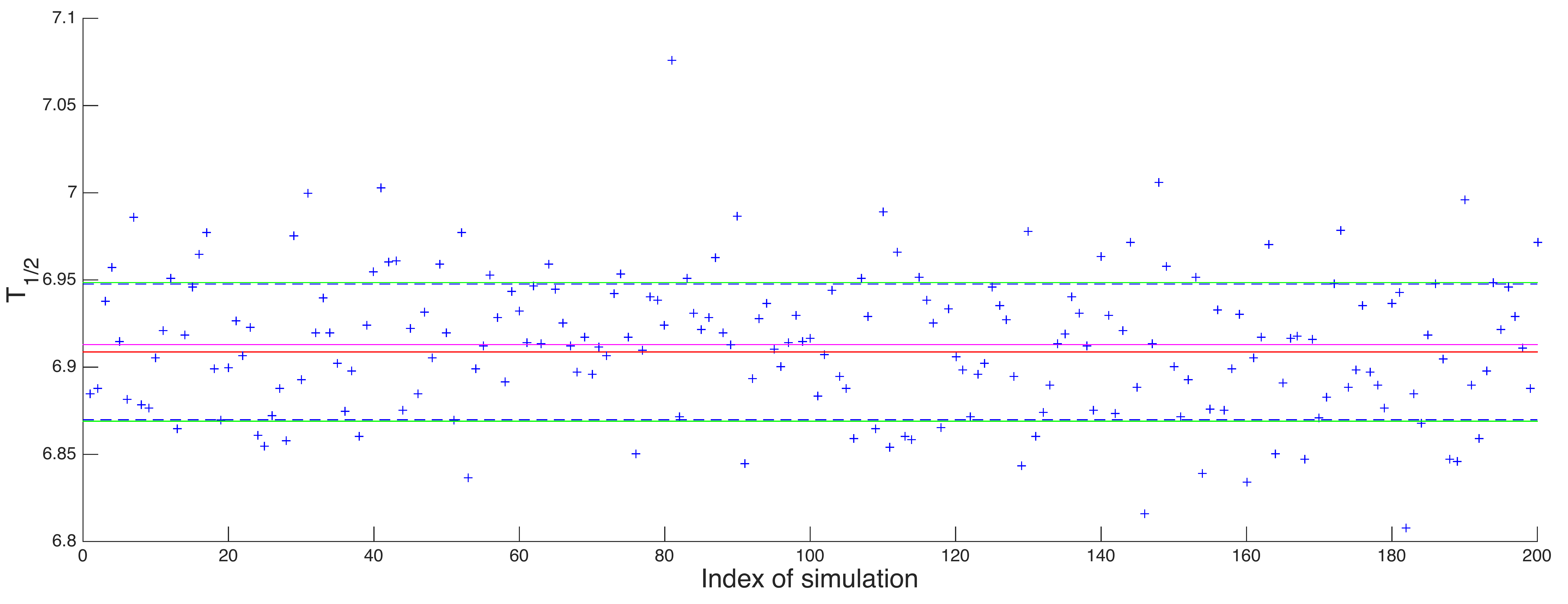}\hss}
\caption{Comparison between predicted mean and standard deviation with simulations. We carried out 200 simulations. For each simulation, $T_{1/2}$ is plotted (blue crosses). The predicted mean and the predicted standard deviation of $T_{1/2}$(red line and green lines), and the simulated mean and simulated variance (pink line and dashed line) are also shown with parameters $M_V=10^{6}$, $\alpha=10^{-3}$, $\beta=1$ and $m=1$.}
\label{suppl:Thalf1}
\end{figure}

\subsection{Qualitative Analysis of the Behaviour of $\mathbf{\bar{x}}$ and $\mathbf{U}$}
\label{suppl:analy}
\subsection*{Behaviour of $\mathbf{\bar x(t)}$}
Recall that
\[
\f{d\bar x}{dt}=-\alpha \bar {x}^2 -\beta \bar{x} (m-\bar{x}) \text{ with } \bar{x}(0)=m, \text{ i.e. } \bar x(t)=m\f{\beta}{\alpha}\f{1}{e^{\beta m t} - 1 +{\beta}/{\alpha}}
\]
\begin{figure}[hbtp]
\centering
{\includegraphics[scale=0.5]{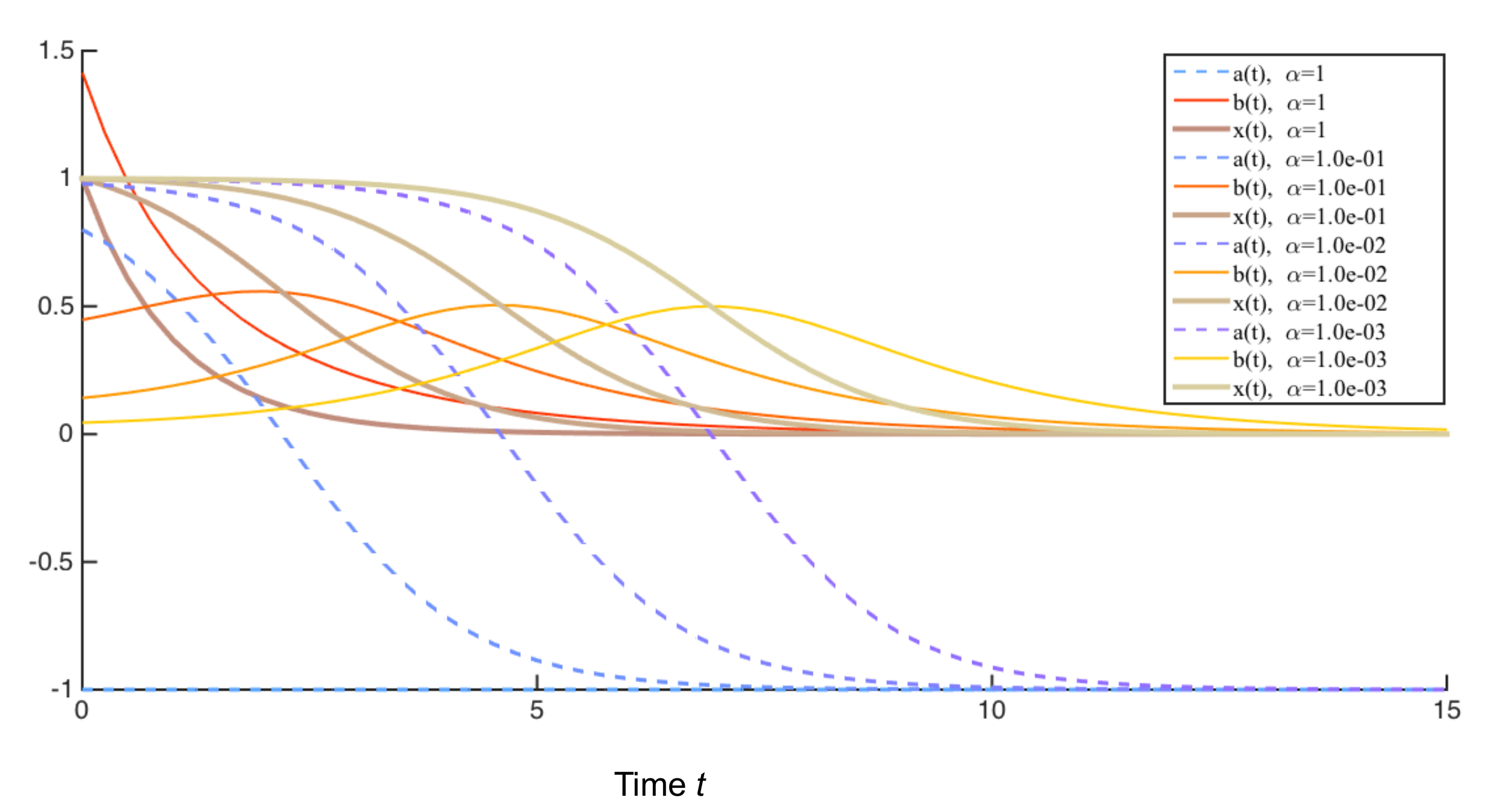}}\par
\caption{Functions $\bar x(t)$  $a(t)$ and $b(t)$    for $\alpha=10^{-k}$, $k=0,\dots, 3$. The function $a(t)$ corresponds to the deterministic part of the equation whereas $b(t)$ corresponds to the stochastic part, see Equations~\eqref{eqaux4} and~\eqref{ab}.}
\label{suppl:fig}
\end{figure}
The limit which interests us is when $\alpha \ll \beta:$ otherwise, the slope at zero, given by $\alpha m^2$, is not small compared to the slope at $t_{1/2}$, which is ${m^2}(\alpha + {\beta})/4$, so that  there is no lag-time, contrarily to what is observed even for high concentrations. In this limit, the formulae for $t_{1/2}$ and $k$ are
\[
t_{1/2}=\frac{1}{\beta m}\log\left(1+{\beta}/{\alpha}\right)\approx\f{1}{\beta m}\log({\beta}/{\alpha})\quad\text{ and }\quad
k=\frac{m\beta}{4}\left(1+{\alpha}/{\beta}\right) \approx \f{m\beta}{4},
\]
so that $\beta=4k/m$ and $\alpha=\beta \exp(-\beta m t_{1/2})=4k \exp(-4kt_{1/2})/m$.

The slope $k$ being measured with little variance between curves of a given concentration, the estimation for $\beta$ is good, at least for a given concentration. What is remarkable is its goodness through different concentrations: our model thus predicts a linear dependence between $k$ and $m$. Concerning $\alpha$, it may change by a typical factor of $\exp( \pm\beta m \sigma)$, so that taking the experimental values of Table~\ref{tab:1} (first, third and fourth columns) we obtain an uncertainty for $\alpha$ which ranges between $7$ and $2.10^5$ according to the set of experiments. This high uncertainty in the estimation of $\alpha$ may to a large extent explain the high variability obtained in the estimated $\alpha$ (See Table~\ref{tab:1}, second column). Note also that this uncertainty does not decrease when the initial concentration increases.

\subsection*{Behaviour of $\mathbf{U(t)}$}
In Figure~\ref{suppl:fig},  the functions $(a(t))$ and $(b(t))$ are plotted  for  fixed $\beta=1$, $V=10^5$ and $m=1$, and various values of $\alpha$ are considered.  
These functions, defined by Equations~\eqref{ab},  drive the dynamics of $(U(t))$ by Relation~\eqref{eqaux4}, 
\[
\diff U(t) =  b(t) \diff W(t) +  a(t)U(t) \diff t.
\]
In particular the coefficient $b(t)$ of the Brownian motion $(W(t))$ modulates the  stochasticity of $(U(t))$.

We observe that for sufficiently small $\alpha:$
\begin{itemize}
\item $a(t)$ begins at $1$, decreases to $-1$. The curves are translations from one another and the time when $a(t)=0$ increases when $\alpha$ decreases.
\item $b(t)$ is nonnegative, bell-shaped, vanishes at zero and infinity,  the curves are translation from one another and its maximum is always the same, around $0.55$. The time at which $b(t)$ is maximum increases when $\alpha$ decreases.
\item At the crossing time, $a(t)=b(t)$ values a constant, around $0.4$ (while $b(t)$ is increasing).
\item The smaller the ratio ${\alpha}/{\beta}$, the higher the average peak value for $\vert U\vert$, and the less noisy each path is (see Figure~\ref{fig:U} for an illustration).
\end{itemize}
All these facts may be deduced analytically from the approximation values when $\alpha \ll \beta$. Denoting $\varepsilon={\alpha}/{\beta}$
\[
b(t)\approx\frac{ \sqrt{ \ep\beta } \sqrt{ e^{\beta m t } +1}}{1+\ep e^{\beta m t} }, \qquad a(t)\approx\beta m\frac{1 -\ep e^{\beta m t }}{1 +\ep e^{\beta m t }}.
\]
For $t=0$, we have $a(0)\approx \beta m=1$, $a(t)$ is clearly decreasing and for $t$ large we have $a(t)\to -\beta m$. This implies that $a(t)$, which has the deterministic influence, leads to exponential growth for $U$ around $0$ and exponential decrease for $U$ around infinity. 

At $t=0$, $b(t)\approx \sqrt{\ep \beta}$ is very small,  $b$ is always positive and at infinity we have $b(t)\approx {\sqrt{\beta}}\exp(-{\beta m t}/{2})/{\sqrt{\ep}}$.

Concerning the crossing point, it occurs when 
$$\frac{ \sqrt{ \ep\beta } \sqrt{ e^{\beta m t } +1}}{1+\ep e^{\beta m t} }\approx\beta m  \frac{1 -\ep e^{\beta m t }}{1 +\ep e^{\beta m t }}.$$
Denoting $d=\ep \exp(\beta m t)$, assuming $\ep\ll \beta m^2$  and taking the square, we have
$d+\ep\approx d\approx \beta m^2(1-d)^2,$
and this gives a value for $d$ which is independent of $\ep$ and $\alpha$, and for this value we have $a{=}b \approx \beta m {(1{-}d)}/{(1{+}d)}$ depending only on $\beta$ and $m$.

This also explains the fact that the maximal value for $\vert U\vert $ increases in average, whereas the "noise" in each path decreases.
These observations are illustrated in Figure~\ref{fig:U}, where we show for each of the previous values of $\alpha=10^{-k}$ with $k=0,\dots,4$ five trajectories for $U_V$ in blue and five trajectories for $U$ in red,  for $M=10^5$. 
\begin{figure}[hbtp]
\includegraphics[scale=0.77]{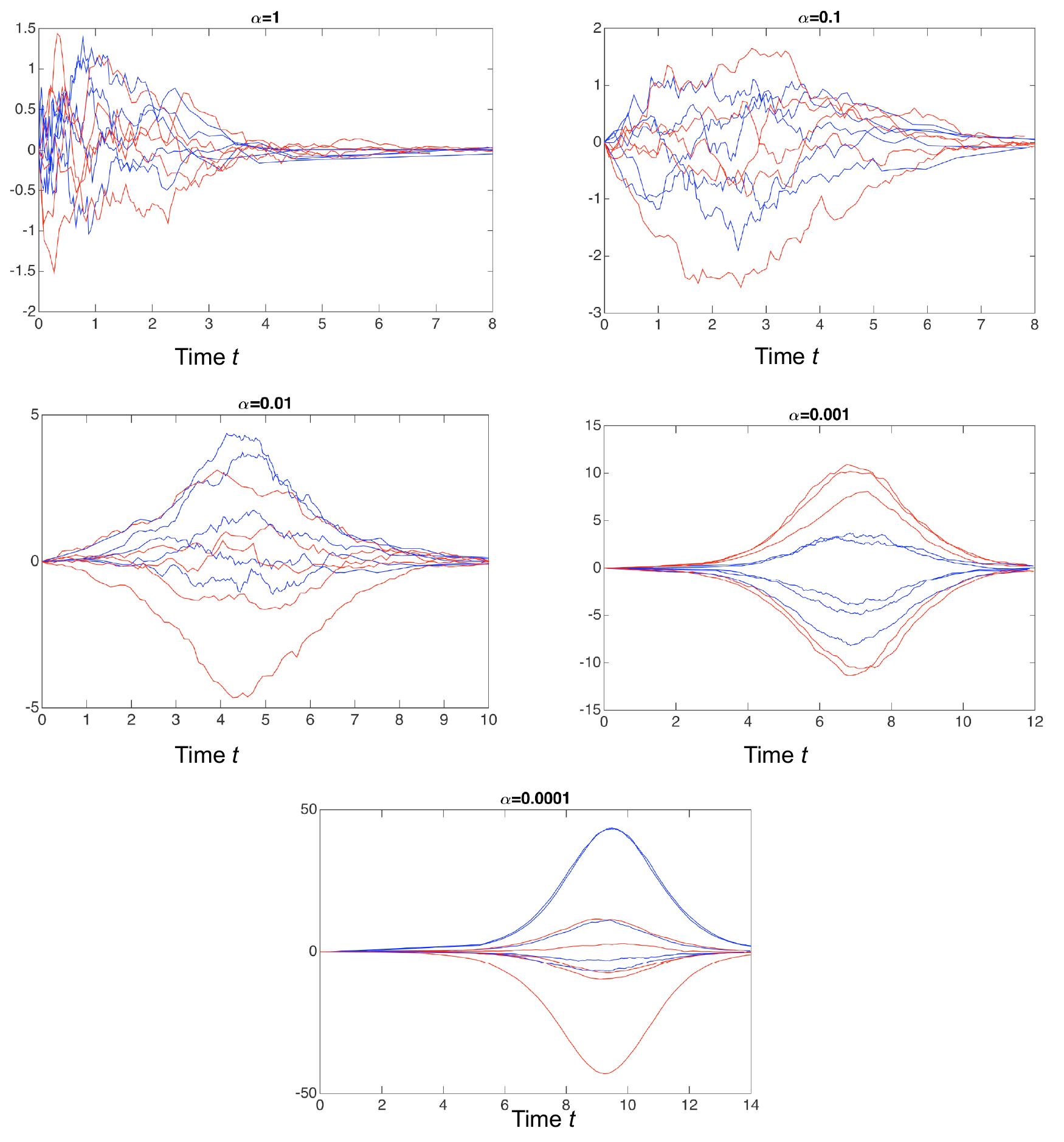}
\caption{Stochasticity of the centered assembly processes $(U_V(t))$ and $(U(t))$. For each of the previous values of $\alpha=10^{-k}$ with $k=0,\dots,4$ five trajectories for $U_V$ in blue and five trajectories for $U$ in red,  for $M=10^5$.  We see that the noise inside each path decreases when $\alpha$ decreases, the stochasticity remaining in the startup of the curves.}		\label{fig:U}
\end{figure}

 \end{widetext}

\end{document}